\documentclass[useAMS,usenatbib,fleqn]{mn2e}

\usepackage{times}
\usepackage[utf8]{inputenc}
\usepackage[T1]{fontenc}

\usepackage{graphicx}
\usepackage{amsmath}
\newcommand{\oj}{OJ~287}

\newcommand{\nmsrs}{3991}

\newcommand{\yr}{\mathrm{yr}}
\newcommand{\dy}{\mathrm{d}}

\newcommand{\sk}{\,}

\newcommand{\redchisq}{\chi^2_\text{red}}

\newcommand{\rsch}{r_\text{S}}

\newcommand{\swwz}{\sigma_\text{WWZ}}

\newcommand{\FAP}{\text{FAP}}

\newcommand{\pbins}{200}
\newcommand{\tbins}{100}

\newcommand{\tuo}{1}
\newcommand{\finca}{2}
\newcommand{\barbados}{3}
\newcommand{\pg}{4}

\usepackage[normalem]{ulem}
\newcommand{\rv}[1]{#1}
\newcommand{\rvd}[1]{}

\title[Short time-scale periodicity in OJ 287]{Short time-scale periodicity in OJ 287}
\author[P. Pihajoki et al.]{
P. Pihajoki$^{\tuo}$\thanks{E-mail: popiha@utu.fi}, 
M. Valtonen$^{\finca,\barbados}$
and
S. Ciprini$^{\pg,\tuo}$
\\
$^{\tuo}$ Department of Physics and Astronomy, University of Turku, 21500 Piikki\"{o}, Finland \\
$^{\finca}$ FINCA, University of Turku, 21500 Piikki\"{o}, Finland \\
$^{\barbados}$ Department of Computer Science, Mathematics and Physics, The University of the West Indies, Cave Hill, Barbados \\
$^{\pg}$ Physics Dept., University of Perugia \& INFN Perugia, 06123 Perugia, Italy  \\
} 

\begin{document}

\date{}

\pagerange{\pageref{firstpage}--\pageref{lastpage}} \pubyear{2013}

\maketitle

\label{firstpage}

\begin{abstract}
We have studied short-term variations of the blazar OJ 287, suspected to host a
supermassive black hole binary. In this study, we use a two-season optical
$R$-band dataset from 2004--2006 which consists of $\nmsrs$ data points from
the \oj{} observation campaign. It has sections of dense time coverage, and is
largely independent from previously published data. We find that this data
confirms the existence of a $\sim50$ day periodic component, presumably related to the
half-period of the innermost stable circular orbit (ISCO) of the primary black
hole. 
\rv{
In addition we find several pseudo-periodic components 
in the 1 to 7 day range, most prominently at 3.5 days,
which are likely Lorentz contracted jet re-emission of the 50 day component.
}
The typical 50 day cycle exhibits a slow rise of brightness and a rapid dimming
before the start of the new cycle. We explain this as being due to a spiral
wave in the accretion disc which feeds the central black hole in this manner.
\end{abstract}


\section{Introduction}

The optical light curve of the blazar \oj{} ($z=0.306$) has been
shown to exhibit periodic double peaked outbursts with a 12-year
cycle \citep{sil1988}. In addition, another significant component
with a period of 60 years has been claimed \citep{val06}. A simple
explanation for both periodicities is a binary black hole system, where
the secondary black hole with an orbital period of 12 years
periodically impacts and perturbs the accretion disc of the primary
black hole. As the secondary is in an eccentric orbit ($e\sim0.7$), the orbit
exhibits strong relativistic precession. 
Therefore the impacts do not occur exactly at equal time intervals, and the
positions of these impact points vary.
This variation can be analysed to constrain the orbital
parameters, which have now been well established \citep{val2010}.

The impacts themselves cause shocked gas to be torn off the accretion
disc of the primary. This gas expands and cools and eventually turns optically
thin, which results in a sharply rising optical flare \citep{leh1996}. These
impact flares along with the tidal flares induced by the perturbation of the
secondary black hole \citep{val2009} explain the 12-year period quite well
\citep{val2011}.
The 60-year period is approximately half of the time it takes for accretion disc
to precess around completely. This period can thus be explained by a
variation in the inclination of the primary disc and a wobble in its jet caused
by the perturbation of the secondary \citep{val2012b}.

Evidence of shorter periodic timescales in \oj{} has also been found
\citep{sag2004,wu2006,gup2012,val2012}, but in general these shorter timescales have been
studied with relatively poorly sampled data in comparison with the well established 12-year and
60-year periods. Shorter timescales can however be significant,
especially in the case of \oj{}, since they convey information of
the inner parts of the primary accretion disc based on the assumption 
that the accretion disc has an inner edge at the innermost
stable circular orbit (ISCO). This then allows an independent measurement
of the corresponding black hole mass, if the spin can be constrained
otherwise. 
In the binary model the impacts of the secondary also happen near the inner
edge of the accretion disc. Therefore it is
possible that this perturbation would result in variations of the
primary accretion rate on rather short timescales. Moreover, at the very shortest
timescales we would expect periodic variations related to the ISCO of
the secondary. Since the secondary may have a high relative spin value, a
non-negligible part of the signal may come from the secondary black hole in the
\oj{} system \citep{pih2013,pih2012}. Thus studies of the short timescales
could give independent constraints on the masses of both black holes.

In this paper we describe observations of the \oj{} from two seasons in
2004--2006. These observations are relatively densely spaced and
allow us to confirm the existence of several periodic components in the
sub-12-year range. 
The data set consists of $\nmsrs$
measurements which are part of an \oj{} monitoring campaign.%
\footnote{The complete list of observers is found in the author list of \citet{cip2007}.}
The data
are largely independent from previously published data and are much more
dense in time coverage than any of the other campaign data. In
particular, coinciding with the \emph{XMM-Newton} pointings in April 2005 and
in November 2005 we had very frequent observations.
While most of the
previous analyses have concentrated on the 12-year periodicity and its
substructure, the shorter time-scale variability has attracted less
attention. To mention a few studies, 
\citet{wu2006}
found a 46 day
periodicity in \oj{} in the spring of 2005. 
\citet{val2012}
confirmed it after nearly doubling the the number of measurements. 

\begin{figure}
\includegraphics[width=\columnwidth]{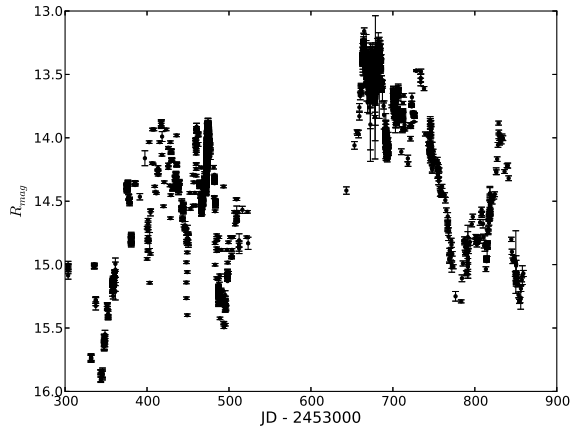}
\caption{\label{fig:alldata}
$R$-band observations of the \oj{} with 1-$\sigma$ error bars.}
\end{figure}

\begin{figure}
\includegraphics[width=\columnwidth]{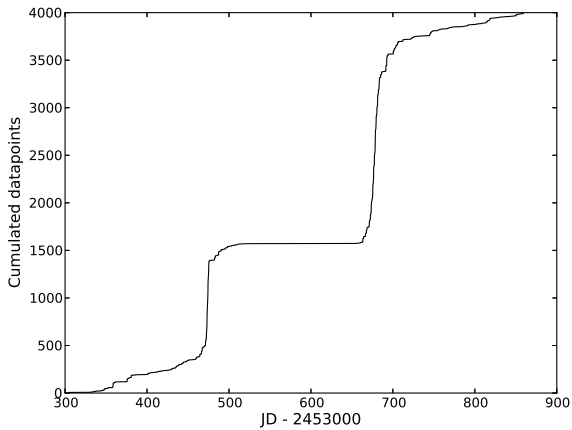}
\caption{\label{fig:data_cdf}
The number of total cumulated datapoints by observation time.
}
\end{figure}

\rv{Here we have a data set of an order of magnitude larger yet,}
and it allows
us to study also a longer interval of time. The data set used in this
paper is illustrated in figure~\ref{fig:alldata}. 
\rv{The rather pathologically uneven sampling of the data is evident from
figure~\ref{fig:data_cdf},
which plots the number of cumulated datapoints by observation time. The vast
majority of datapoints have been recorded in the space of just a few days,
centered on days 474 and 679 in JD-2453000. To search for periodicities from
data of this character, such methods are needed that carefully account for this
pathology. We will describe these in the next chapter.
}

\citet{gup2012}
have pointed out that periodicities in
extragalactic jets may arise from instabilities of flow at ISCO. 
Thus the period of this orbit may be imprinted in
the light curve structure. 
\citet{val2012}
note that in a binary
system the relevant period is one half of the full period due to bipolar
influence of the tides from the companion. 
\citet{gup2012}
also
mention that the same periodic structure may be imprinted in the jet
itself where it becomes shortened by the Lorentz factor of the
relativistic flow. However, we would not expect more \rv{than} a preferred
variability time-scale from this process, as the orbital phase would be
lost.  

\section{Periodicity search}

\rv{
To look for possible periodicities and ascertain their significances, we employ
two methods: the Lomb--Scargle periodogram (LSP) \citep{sca1982}
and the Weighted wavelet Z-transform (WWZ)
\citep{fos1996}. In addition we also experiment with data binning to
even out the effect to the two extremly dense observation regions. For this purpose,
we produced a dataset with data in figure~\ref{fig:alldata} divided into 1 day bins.
}

\begin{figure}
\includegraphics[width=\columnwidth]{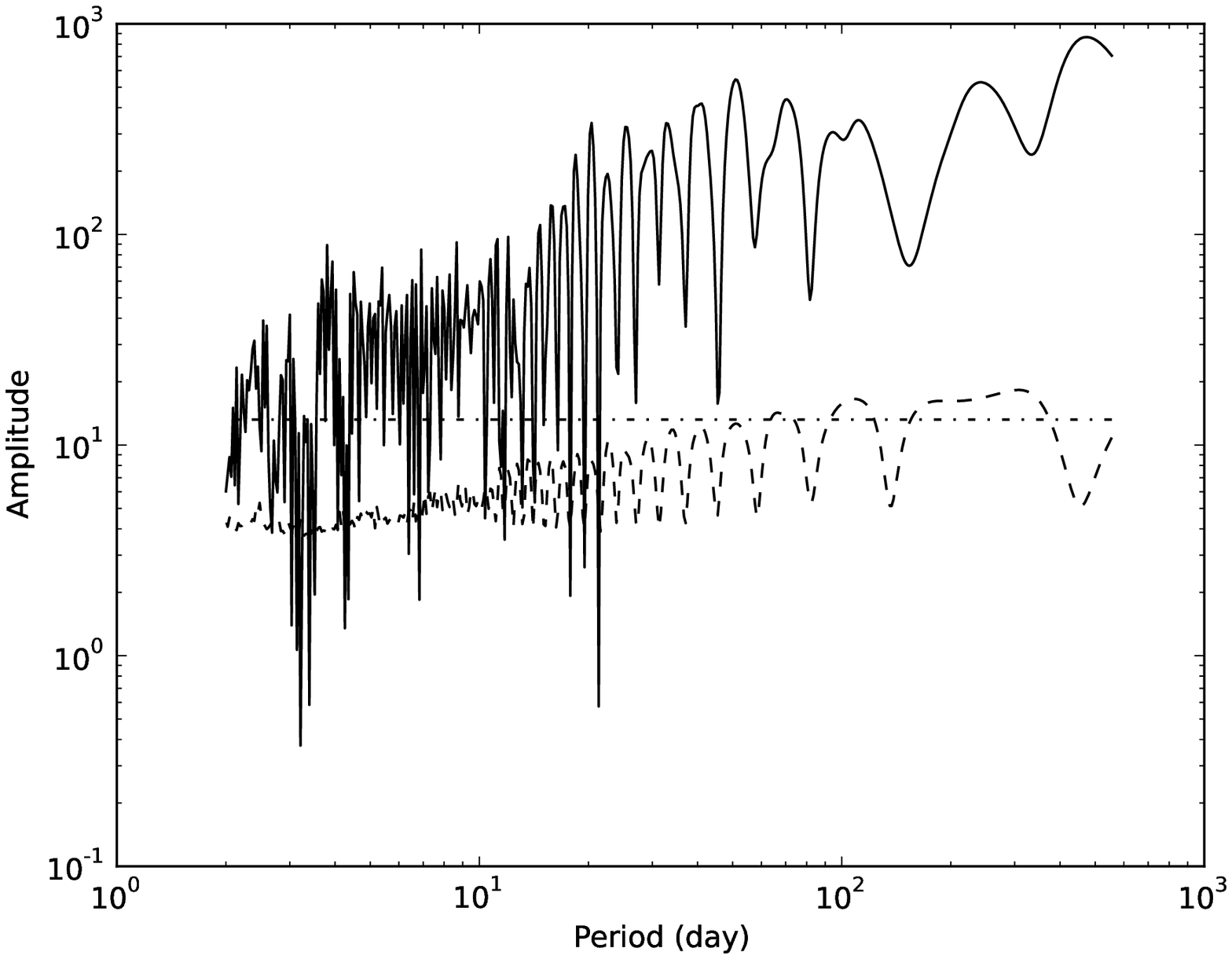}
\includegraphics[width=\columnwidth]{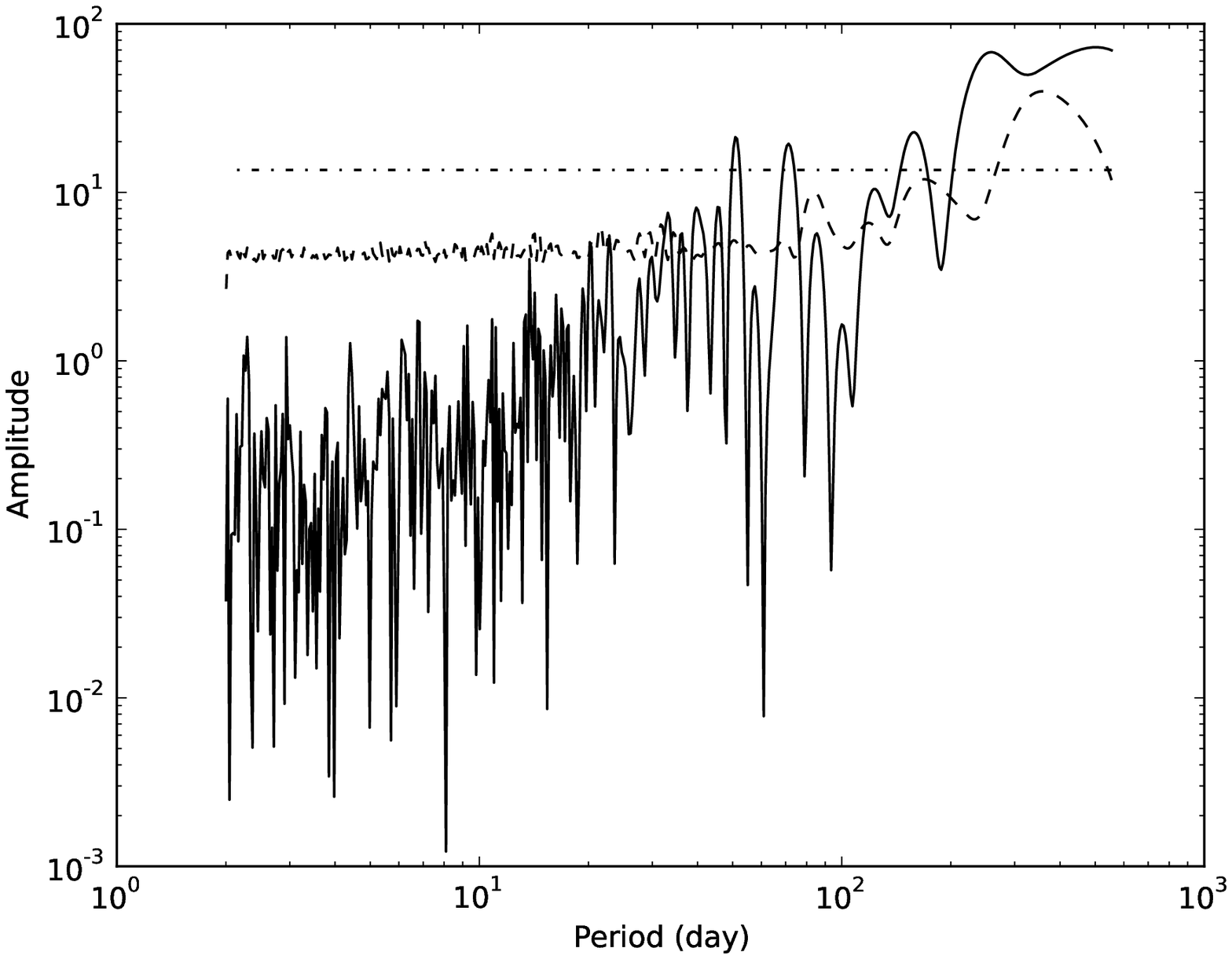}
\caption{\label{fig:alldata_lsp}
Lomb--Scargle periodogram of the entire dataset (top),
\rv{
and of the 1 day binned data (bottom). 
Periodogram is drawn with solid line, bootstrapped $3\sigma$ significance limits
with dashed line and \citet{bal2008} analytic significance limit for $p=0.0014$ with
dash-dotted line.
}
}
\end{figure}

\rv{
The first method, LSP, is temporally global in the sense that it presents
information of possible periodicites by considering the entire data all at
once.
The method attempts to compensate for the uneven spacing of data, a serious
problem in many astronomical time series, and particularly so in the case of OJ
287. The LSP does this by projecting the data on trial functions, $\sin(\omega
t)$ and $\cos(\omega t)$, where $\omega$ is the angular frequency,
essentially resulting in date compensation as in \citet{fer1981}.
}
\rv{
We calculated the Lomb--Scargle periodograms for the entire dataset and the
1-day binned dataset with the results illustrated in
figure~\ref{fig:alldata_lsp}. The periodograms were calculated at 500 points
from $f_\text{min} = 1/T \sim 1.83\cdot10^{-3}\sk\dy^{-1}$ to $f_\text{max} =
0.5\sk\dy^{-1}$, distributed logarithmically, with $T=555.533813\sk\dy$ the
time extent of the total data set.
}

\rv{
To quantify the significance of the peaks in the LSP, we used a bootstrapping
procedure \citep{fei2012}. In this procedure we resampled the original data
1000 times, keeping the observation times constant.  Periodograms were then
calculated from the resampled data, and the mean $\mu_\text{LSP}$ and sample
standard deviation $\sigma_\text{LSP}$ for each period were calculated from
these. As the limit of significance, we set
$\mu_\text{LSP} + 3\sigma_\text{LSP}$, i.e. $3\sigma$. In addition to this
numerical limit, we also employed the analytic maximum bound for the LSP false
alarm probability found by \citet{bal2008},
\begin{equation}
\FAP(z) \approx W\exp(-z)\sqrt{z},
\end{equation}
where $W = f_\text{max}T_\text{eff}$, $T_\text{eff} = \sqrt{4\pi\sigma_t^2}$,
$\sigma_t^2$ is the weighted variance of the
observation times and $f_\text{max}$ is the maximum frequency for which the
periodogram has been calculated.
}

\rv{
From figure~\ref{fig:alldata_lsp} it is immediately evident that the
LSP for the entire dataset seems to rise above the significance limit
almost everywhere on the observed frequency spectrum. We are confident
that this effect is 
due to the extremely pathological time sampling. As is evident from figure~\ref{fig:data_cdf}, the
original data represents essentially two point samples with unduly high weights
i.e. the highly sampled regions, with the rest of the data scattered with more
consistent spacing. Observing the $3\sigma$ limits from bootstrapped
data strengthens this doubt, as the peaks and throughs of the
bootstrapped LSPs align well with the peaks and throughs of the original
LSP, indicating that the uneven timing is contributing significantly to
them. Thus we affix some doubt on the LSP results of the whole dataset.
}

\rv{
We next focus on the more well behaved binned data. From
figure~\ref{fig:alldata_lsp} we find several interesting periodicities
exceeding both our numerical and analytical $3\sigma$ significance
limits. The most prominent components have long periods of $\sim 500$, 
$\sim 260$ days and $\sim 150$ days.
From the bootstrapped $3\sigma$ curve, we see that the detected 500 and 150 day
periodicities may arise partly due to effects of the data sampling.
Furthermore, the length of the 500 day period is near that of the total
length of the data set, casting further doubt on its significance. The 260 day
period however is apparently a genuine detection. At higher frequencies,
there are two significant peaks at $\sim 70$ and $\sim 50$ days. We will
examine these three significant periods further with the Weighted
Wavelet Z-transform method.
}

\rv{
The WWZ method in constrast is local, in the sense that it gives information on
periodicity near some specific point of time in the data.
The method, like LSP, projects the data, but using $\cos[\omega(t-\tau)]$, 
$\sin[\omega(t-\tau)]$ and the constant function $\mathbf{1}(t) = 1$. The projection in
addition uses weights of the form $\exp(-c\omega^2(t-\tau)^2)$, with $c$ a
tunable parameter. In essence the method examines data for periodicities
near time $\tau$, in a neighbourhood defined by $\omega$ and $c$.
In our case, we adopt two choices. First, $c=0.0125$, as originally proposed in
\citet{fos1996}, for improved time resolution on shorter segments of the data.
Second, $c=0.005$, used for the entire dataset to improve frequency resolution.
This translates to the wavelet decaying by $e^{-1}$ in $1/(2\pi\sqrt{c}) \sim
1.4$ cycles in the first case and $\sim 2.4$ cycles in the second case.  The
values can be compared to e.g. $c=0.001$ used in \citet{tem2005} and
\citet{you2012}, for
longer datasets than the one used here. To reduce the low power noise in the data, we also
detrended the data with a linear fit before calculating the WWZ statistic.
}

\rv{
To quantify the errors in the detected periodicities, and their significances,
we employed a two-fold approach following \citet{you2012}. 
We used bootstrapping to establish the significance of period, by resampling the
calculated WWZ statistic 1000 times, and calculated the standard deviation of
the entire statistic for each resample. From this we obtained an estimate for the
standard deviation of the WWZ statistic, $\swwz$. We then chose to
use $3\swwz$ as our criterion for detection, thus setting the one-tailed 
$p$-value to $p\sim0.0014$. To estimate the \emph{maximum} error in a detected period,
we used the confusion limit estimation method of \citet{tem2005}.
In this method, the half width half maximum (HWHM) of the peak in the WWZ statistic is
used as the maximum $1\sigma$ uncertainty of the period. 
}

\begin{figure}
\includegraphics[width=\columnwidth]{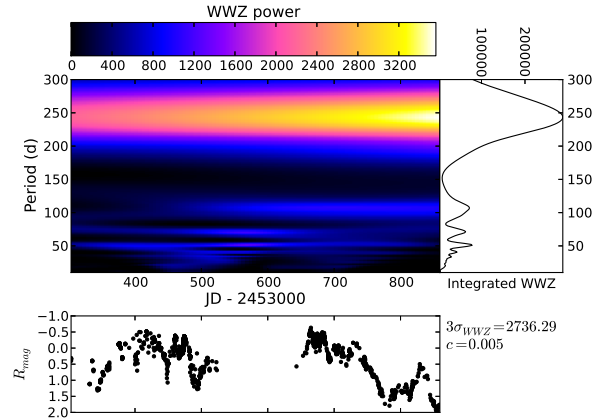}
\caption{\label{fig:wwz_alldata}
\rv{
The WWZ statistic of the raw data (top left), using $\pbins$ period bins and
$\tbins$ time bins. The integrated WWZ power (right). Data points used (bottom).
}
}
\end{figure}

\begin{figure}
\includegraphics[width=\columnwidth]{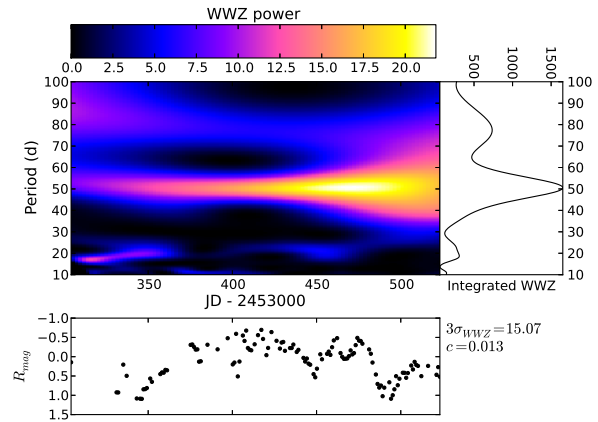}
\caption{\label{fig:wwz_1st_half}
\rv{
The WWZ statistic of the first half of the data binned in 1 day bins (top left), using 
$\pbins$ period bins and $\tbins$ time bins. The integrated WWZ power (right). Data points
used (bottom).
}
}
\end{figure}

\begin{figure}
\includegraphics[width=\columnwidth]{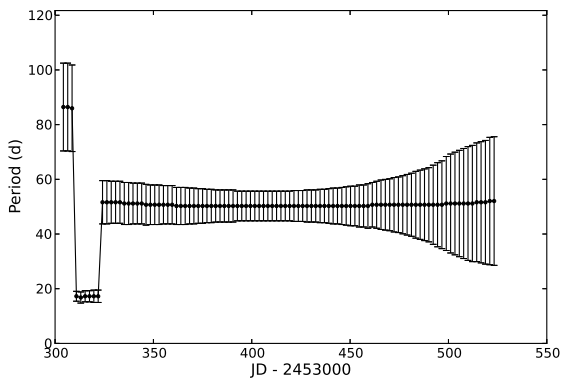}
\caption{\label{fig:wwz_peri_1st_half}
\rv{
The dominant period extracted from figure~\ref{fig:wwz_1st_half}. Error bars
indicate maximum $1\sigma$ errors calculated by the confusion limit method.
}
}
\end{figure}

\rv{
The WWZ statistic for the entire data is plotted in figure~\ref{fig:wwz_alldata}, using
$\tbins$ time bins and $\pbins$ period bins of equal size, spanning periods from 10 to 300 days. 
The upper limit is slightly over half of the length of data, which limits our
search to periodicities that have at least $\sim2$ complete cycles.
}

\rv{
We find a dominant periodicity of $\sim250$ days above the $3\swwz$ level,
continuing throughout the data. 
Two of its higher harmonics seem to appear at $\sim120$ and $\sim70$ days.
However, these detections are not significant, and furthermore their peaks are
located in or adjacent to areas where there are no datapoints, which leads
us to surmise that they are probably spurious.
}

\rv{
We repeated this calculation for data binned in 1 day bins
to see whether this mitigation of sampling
effects would change our conclusions.
The results 
show that at least for our choice
of $c$ the conclusions are not affected, though the detected maximum
peak shifts to about 260 days.
}

\rv{
In both cases, we do find an additional peak at $\sim50$ days located in the
first half of the data. The peak however is not detected with significance, and
is located close to the edge of the data gap in the middle of the data. Thus, we
will investigate it further. To this end, we separated the two observing seasons
from the one day binned data, and calculated the WWZ statistic for both parts.
In this case we searched for periods from 10 to 100 days, with the same
rationale as above. For these shorter segments of data, we set $c=0.0125$. The
WWZ statistic for the first half is shown in figure
\ref{fig:wwz_1st_half}, with the dominant period along with error bars
is shown in figure \ref{fig:wwz_peri_1st_half}. 
}

\rv{
The WWZ statistic for the first half of the data
shows a significant detection of a $\sim50$ day period, though the
significance does not persist throughout the data length.
This is confirmed by the results of the latter half
where the 50 period is present but no longer significant. 
Instead we find signicant detections of 
a period trending from 70 to 80 days at the very end of the second
observing season.
Within the limits of uncertainties, it is possible that this could be a
higher harmonic of of the 250 day period.
}

\rv{
We finally extend the search using the WWZ method to periods under 10 days. 
For these purposes, there are two usable segments in the data, centered around
474 and 679 in JD-2453000. 
}

\begin{figure}
\includegraphics[width=\columnwidth]{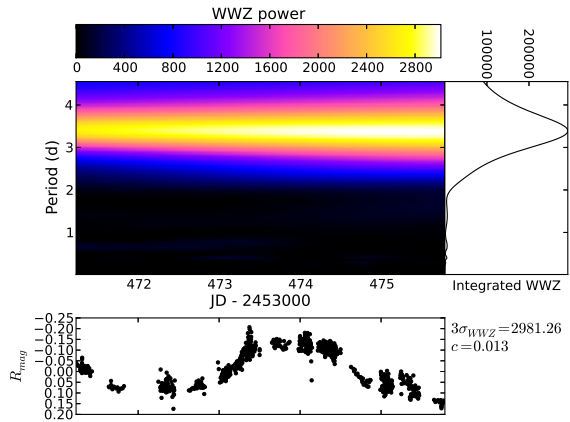}
\caption{\label{fig:wwz_474}
\rv{
The WWZ statistic of data around day 474 (top left), using $\pbins$
period bins and $\tbins$ time bins. The integrated WWZ power (right).
Data points used (bottom).
}
}
\end{figure}

\rv{
We calculated the WWZ statistics for these two subsets of the data, using the
same number of bins as before, but using $c=0.0125$ and spanning periods from
0.01 days to 4.55 days for data in around day 474 
and 0.01 days to 21.5
days for data around day 679. 
The maximum period is thus equal to the full extent of the data,
relaxing our previous search constraint now
to periodicities with \emph{one} complete cycle in the data, with the caveat that even
a significant detection is not enough to rule out the chance of observing such a periodicity
by accident.
The results for the data around day 474 are shown in 
figure \ref{fig:wwz_474}. 
}

\rv{
We find a significant detection of a $\sim3.5$ day period in the first dataset
(figure~\ref{fig:wwz_474}). The detection is not significant throughout the
entire data slice, and the period is near the length of the dataset. Thus we
must classify the detection as rather tentative, and requiring verification with
future data.
}

\rv{
The second dataset is dominated by a $\sim20$ day period, equal to the length in
the data, with an additional periodic component in the 7 day range. Both of
these peaks rise above the significance level near the end of the dataset,
though barely, and as the length of the data is 21.5 days, the 20 day
period is likely spurious.
Both periodicities are then also tentative detections at best,
and underline the need for high time resolution observations of \oj{}
extending for months rather than days.
}

\rv{
Finally, if we divide the second dataset further into two halves at day
676.5, both halves do show the 3.5 day sinusoidal variation, but the
phases do not match at the border (figure~\ref{fig:short}). Thus the
3.5 day feature lives for one period at best.}

\section{Discussion}

\rv{We now have statistically significant detections of periods of
approximately 250 and 50 days to consider. In addition, there is
a 3.5 day variability time-scale and a borderline detection of a
$\sim70$ day periodicity.
}

The baseline over two years is well modeled by \rv{the detected}
periodic component with about $250\sk\dy \sim 0.7\sk\yr$ period. The
time coverage of this data set is too short to say much about this time
scale. In the binary model of \citet{leh1996} and \citet{sun1997}
the half-period of the disc at about
9 Schwarzschild radii of the primary black hole has this value. 
As the impacts of the secondary on the disc happen just
outside this radius, the accretion flow variations may well have this
time-scale (see e.g. \citet{val2009} for the structure of induced flows).
\rv{
The induced brightness variations have no a-priori reason to be perfectly
sinusoidal.  This is supported by the fact that in addition to the 250 day
baseline, WWZ peaks near harmonic periods of 130 and 70 days were
also detected. Thus the 70 day component is possibly a harmonic
component of the 250 day base period.
}

The second \rv{independent} component that shows up significantly is
about 50 day periodic variation.
\rv{
The feature is significanctly present in the first observing season, but
not so in the second. It could be recurrent in nature, but data from
more observing seasons would be required to settle this.
}
The
exact value of the period is not well determined but it is probably not
very far from $50\pm5$ days.
The result is quite robust and independent
of the exact data set used, and whether the 2005 October outburst data
are included or not. Since in the binary model the latter outburst is
not related to the jet 
\citep{val2012b},
it may be better to eliminate this burst from the analysis, either by
deducting the standard outburst light curve \citep{val2011b}
from the observed data, or
by excluding a short section of the data from the analysis. 
\rv{
However, in this work, we found that the conclusions were not affected
by the deduction of the outburst curve, and thus the unsubtracted
original data were used throughout, except where explicitly noted.
}

\begin{figure}
\includegraphics[width=\columnwidth]{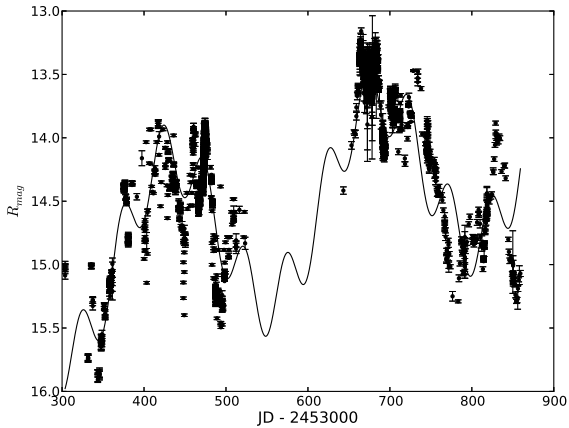}
\caption{\label{fig:slowperiods}
\rv{Raw data and a
fit with periodic components of 49.9 days and
251 days plus a linear trend.
}
}
\end{figure}

\rv{Both the $250\sk\dy$ and $50\sk\dy$ periods} are illustrated in
figure~\ref{fig:slowperiods}, \rv{where we have fit a linear combination of the
sinusoids together with a linear trend. The resulting parameter values and
$3\sigma$ uncertainties along with reduced chi-square $\redchisq$ and
coefficient of determination $r^2$ are listed in
table~\ref{tb:periodresults}.}
In calculating the $\redchisq$ values we
accounted for the inherent noisiness of the source by first dividing the data
slice used into bins of length $P/3$ where $P$ was the shortest period under
consideration. We then used the variance calculated from each bin as the
variance of the data points of that bin in the calculation.
\rv{
Despite the WWZ detections, the calculated $\redchisq$ value for the fit
in figure~\ref{fig:slowperiods} is rather low. This is not surprising,
considering that the 50 day period is much lower in amplitude during the
second observing season, where the 70 day period dominates.
}

The Lorentz gamma factor of \oj{} jet is about 14 \citep{hov2009}.
Thus another interesting time-scale would be the observed 50 day period
divided by the Lorentz factor, i.e. 
$\sim3.5$ days. 
\rv{This period shows up strongly during the first intense observation period
(figure~\ref{fig:wwz_474}), but is not statistically significant during the
second intense observing period. This is not surprising,
since jet re-emission can probably be expected to be quasi-periodic at best.
In the latter half a periodicity of 7 days was marginally detected. 
This obviously results from the 3.5 day sinusoidal variations in the two
segments of the second intense campaign with a change of phase in the
middle of it.
}

\begin{figure}
\includegraphics[angle=-90,width=0.8\columnwidth]{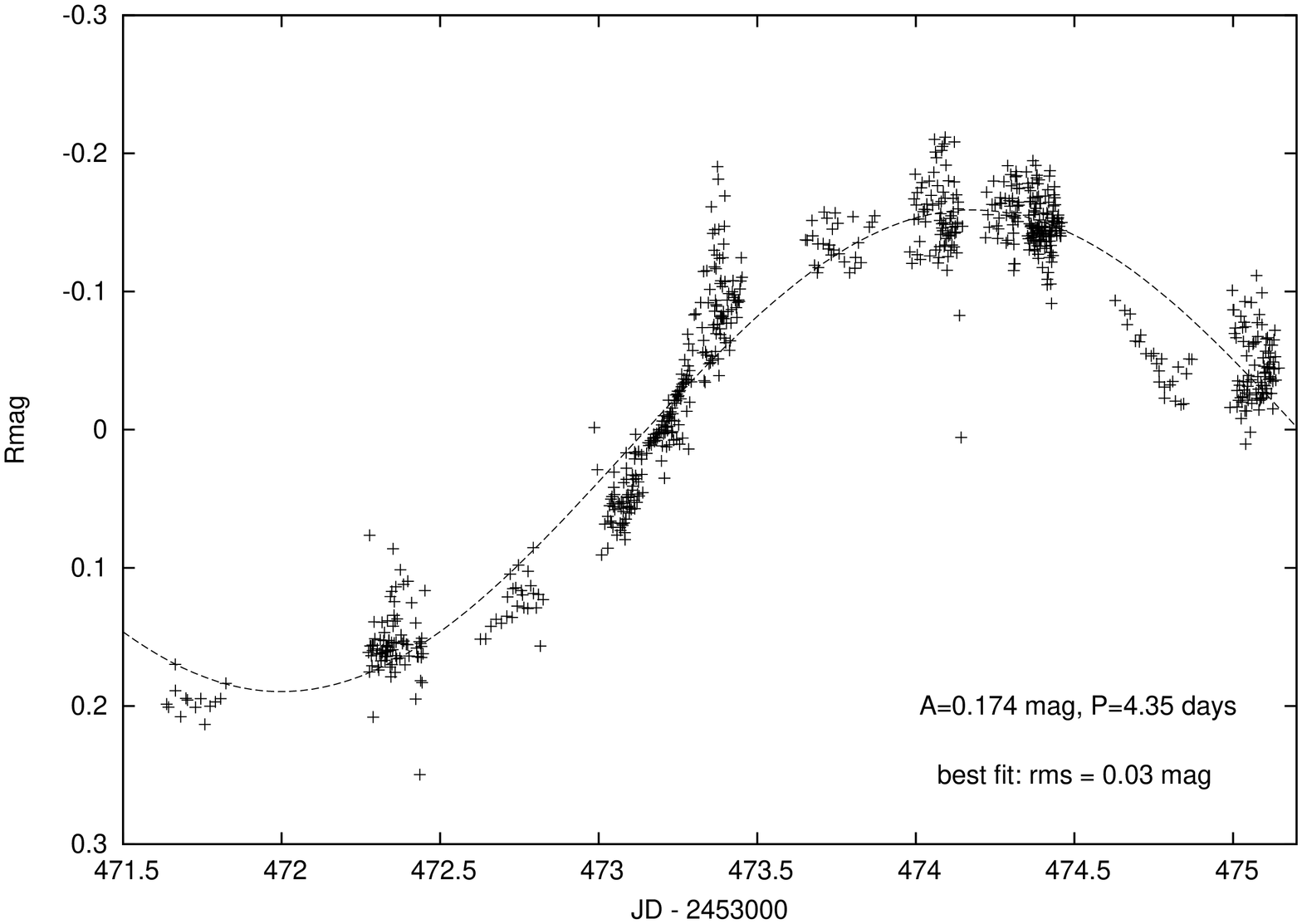}
\includegraphics[angle=-90,width=0.8\columnwidth]{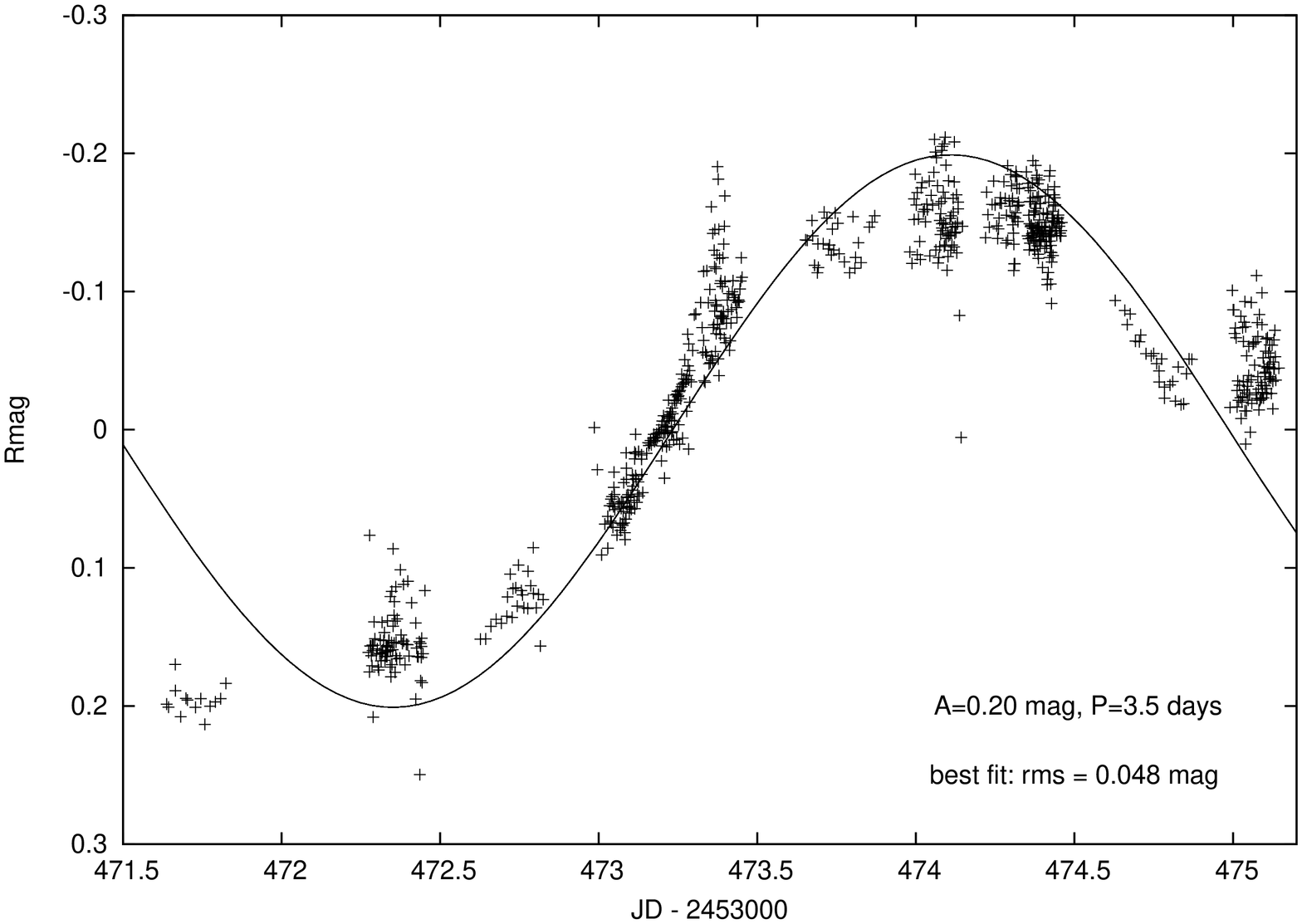}
\includegraphics[angle=-90,width=0.8\columnwidth]{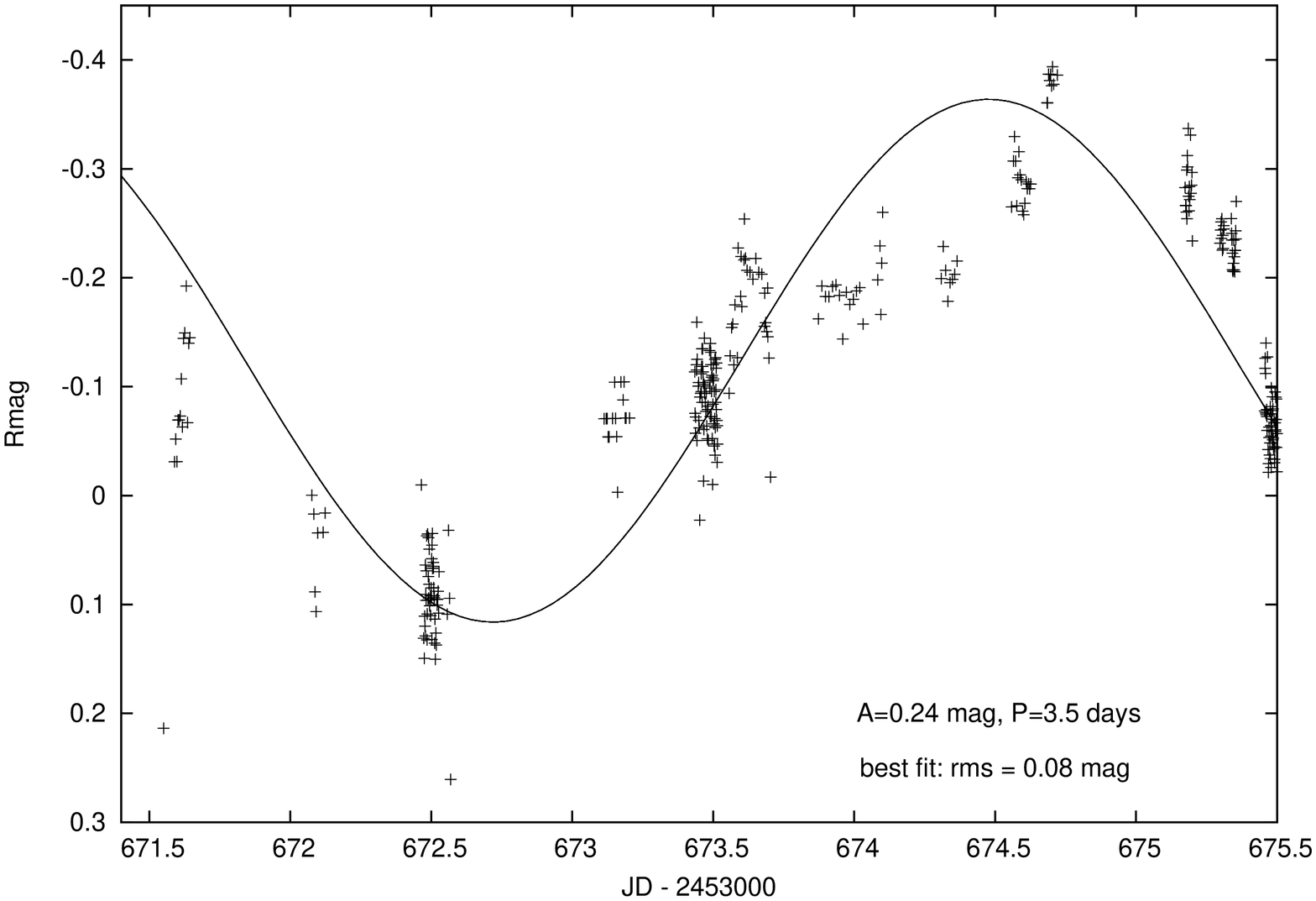}
\includegraphics[angle=-90,width=0.8\columnwidth]{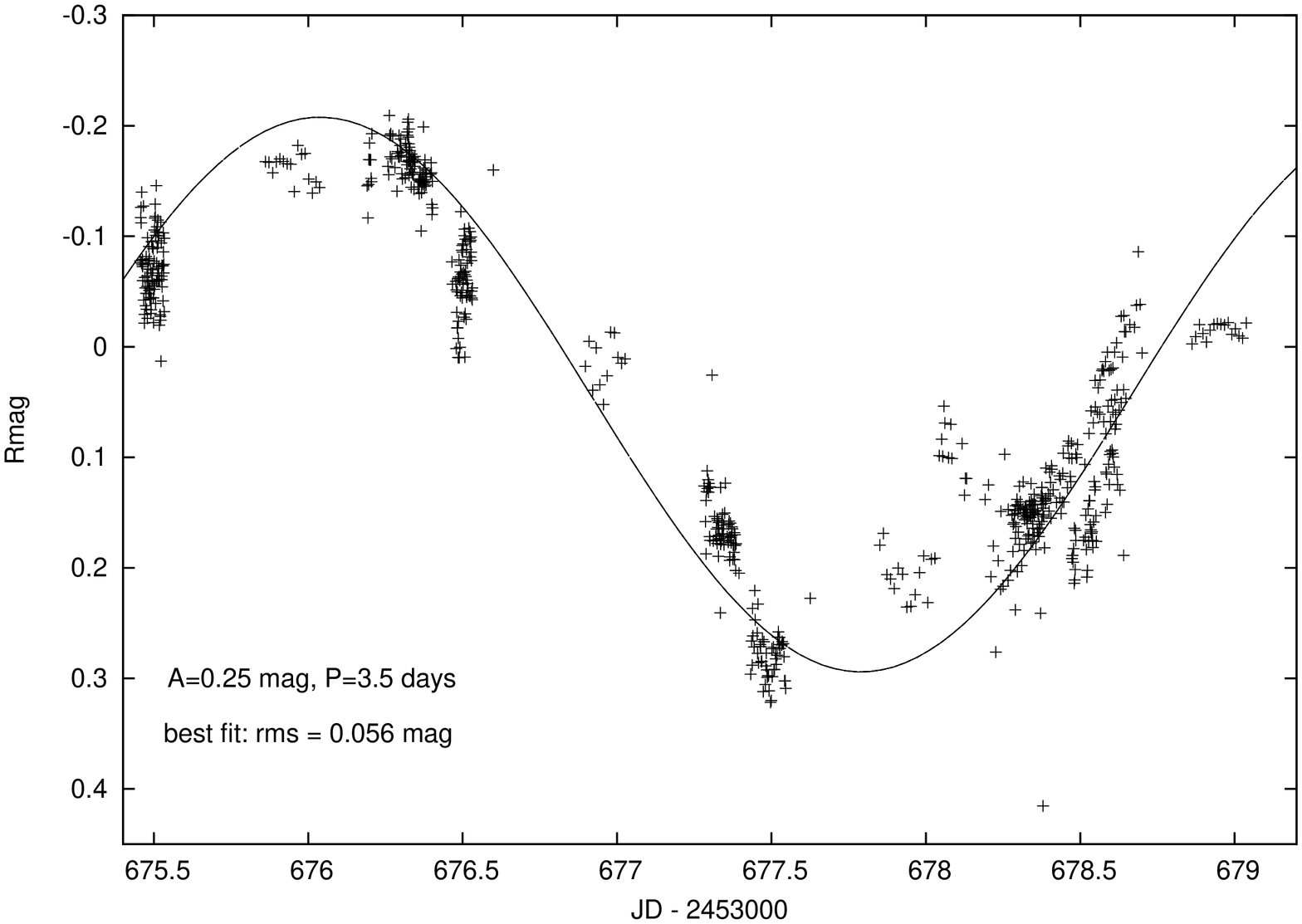}
\vspace{1cm}
\caption{\label{fig:short}
Slices of the dataset illustrating the
Lorentz contracted short period fits of 4.35 days (\rv{topmost}) and 3.5 days (below).
}
\end{figure}

\rv{
Several periodic least squares fits of how the 3.5--4.5 day periods do
emerge from the data are shown in figure~\ref{fig:short}. As the data
slices only span approximately one period of each, they must be
considered illustrative at this point. Parameters from these fits are
displayed in table~\ref{tb:periodresults}.
}
No statistically significant shorter periods were found.

Theoretically the lack of the shorter components is understood, since the
secondary was behind the disc of the primary during both of the intense
campaigns. The frequency of observations prior to February 2005, when the
secondary could have been visible, was too low to search for very rapid
variations in the time-scale of hours.

\begin{table*}
\begin{center}
\caption{\label{tb:periodresults}Found periodicities. Parameters given with
3-$\sigma$ errors, rounded up. Values $a$ and $b$ refer to the linear fit $y =
a(t-t_0) + b$ if used.}
\begin{tabular}{@{}lrrrrrrr@{}}
\hline
Data slice (JD-2453000) & Period ($\dy$) & Amplitude (mag) & a & b & RMS error (mag) & $\chi_\text{red}^2$ & $r^2$ \\
\hline
\multicolumn{8}{l}{$\geq 50\sk\dy$ periods as one fit} \\
303.716187--859.250000  &   251 $\pm$ 2   & 0.68 $\pm$ 0.03 & & & & & \\
                        &  49.9 $\pm$ 0.2 & 0.27 $\pm$ 0.02 & & & & & \\
                        &                 &                 & $(-1.87\pm0.09)\cdot10^{-3}$ & $15.07\pm0.03$ & \multicolumn{3}{r}{given for the linear combination} \\
	                &                 &                   & & & 0.21   & 4.9   & 0.85 \\
\multicolumn{8}{l}{$<50\sk\dy$ periods, with 2005 peak and above periods removed} \\
471.637939--475.144318  & 3.88 $\pm$ 0.09 & 0.16  $\pm$ 0.01  & & & 0.037  & 0.41  & 0.89 \\
671.550415--675.499817  & 3.8  $\pm$ 0.2  & 0.18  $\pm$ 0.03  & & & 0.076  & 1.04  & 0.70 \\
675.456116--679.037292  & 3.51 $\pm$ 0.06 & 0.211 $\pm$ 0.007 & & & 0.051  & 0.42  & 0.88 \\
\hline
\end{tabular}
\end{center}
\end{table*}

\section{The 50 day period}

\begin{figure}
\includegraphics[width=\columnwidth]{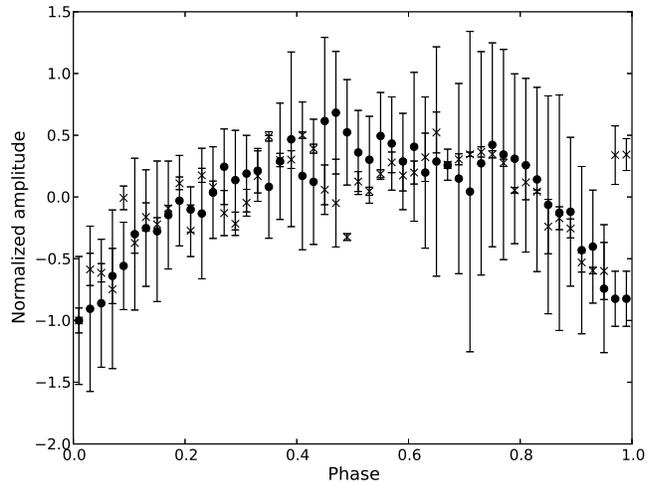}
\caption{\label{fig:phase_diagram}
Data phase wrapped with a 51 day period in 50 bins (points) and
logarithm of simulated accretion counts phase wrapped with a 9.26 year
period in 50 bins (crosses). Normalized bin medians are shown, with
error bars equal to $1\sigma$ derived from MADN.
}
\end{figure}

To further study the 50 day period, we phase wrapped the data from 2004--2005
with a period of 51 days, encompassing six cycles. The result can be seen in
figure~\ref{fig:phase_diagram}. 
\rv{The figure shows the medians $\text{Med}(b_i)$ of each bin $b_i$
with error bars equal to normalized median absolute deviation,
$\text{MADN} = 1.4826\cdot\text{Med}(b_i-\text{Med}(b_i))$, equal to
1-$\sigma$ for normally distributed data \citep{fei2012}.}

The periodic outburst shows a distinguishable
pattern of slow rise to maximum brightness, and subsequent rapid collapse. This
could plausibly result from a spiral density wave pattern in the accretion disc
of the primary black hole.

Such spiral density waves have been reported in magnetohydrodynamical
(MHD) simulations of non-self-gravitating accretion discs
\citep{tag1999,cau2001,haw2001},
self-gravitating particle disc simulations \citep{ant1988},
and purely hydrodynamical simulations of thin accretion discs
\citep{ant1988,li2001}.
Binary black hole configurations also give rise to spiral waves in circumbinary accretion discs
\citep{han2010}.
Thus the accretion disc or discs of the \oj{} system are likely to harbor spiral density waves.

\citet{pih2013} discuss the profile of an outburst originating near
the ISCO of a black hole in a binary system, specifically considering the
\emph{secondary} in the \oj{} system.
They note that both a perturbation by the other body in the system and
the presence of a magnetic field can lead to quick accretion rates near the ISCO 
\citep{byr1986, byr1987, lin1988, goo1993, kro2005}. This result naturally
pertains to the accretion near the ISCO  of the primary black hole as well, with
the time-scales scaled up accordingly.

To investigate this possible origin of the profile of the periodic signal, 
we performed a series of
numerical simulations of the accretion disc of the primary. We used a
particle simulation n-body code explained in detail in \citet{pih2013}.
The simulation was set up with a zero thickness disc of $n=10^5$ 
particles around the primary black hole spanning the radial range of 25 to 50
Schwarzschild radii of the primary. The particles were set up with an $r^{-2}$
density profile, with an initial spiral wave perturbation in the
\emph{velocities}, adapted from the formulae
used by \citet{har2002}
\begin{gather}
v_r = v_k  \epsilon_r  \exp\left(-\frac{r-r_\text{in}}{r_0}\right) 
\cos\left[m\phi - k(r-r_\text{in})\right] \\
v_\phi = v_k\left\{1 + \epsilon_\phi  \exp\left(-\frac{r-r_\text{in}}{r_0}\right) 
\cos\left[m\phi - k(r-r_\text{in})\right]\right\}.
\end{gather}
Here $v_k$ is the Keplerian velocity at distance $r$ from the primary black
hole, $r_\text{in}$ is the inner edge of the disc, $r_0 = 25\rsch$ is the dampening factor, 
$\rsch$ is the primary Schwarzschild radius,
$\epsilon_r = \epsilon_\phi = 0.1$ are the amplitudes of the perturbation and $m=2$,
$k=2\pi/25\sk\rsch$ control the number of arms and tightness of the spiral.
The black hole binary was set up with orbital elements and initial conditions from
\citet{val2010} and other parameters such as artificial viscosity set up
as in \citet{pih2013}, with $\alpha=0.03$.
We then followed the flux of particles through a radius of $r=25\rsch$
around the primary black hole during the course of the simulation. 
This distance was chosen as following the particles near the ISCO of the
primary black hole either needs taxingly low timesteps or induces numerical
error. 
\rv{The half-period orbital timescale of 50 days at the ISCO of the primary black
hole must thus be scaled up to this new artificial ISCO, which
leads to a timescale of 
$50\sk\text{days}\sk\times(25/1.7)^{3/2}\sim 8\sk\yr$}.
The secondary crosses the disc near $10\rsch$ of the primary. 
Therefore counting the accretion at the greater orbital distance than ISCO
produces an amplified accretion rate. In the \oj{} disc we expect the accretion
rate variations at ISCO to be lower in amplitude.
The total length of the simulation was 40 years.

\begin{figure}
\includegraphics[width=\columnwidth]{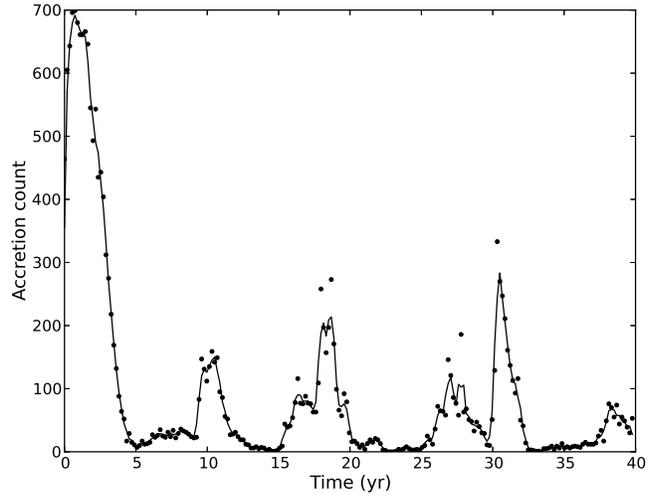}
\caption{\label{fig:accretion}
Accretion count of the primary black hole with a 3 point moving average also
displayed (solid line).}
\end{figure}

\begin{figure}
\includegraphics[width=\columnwidth]{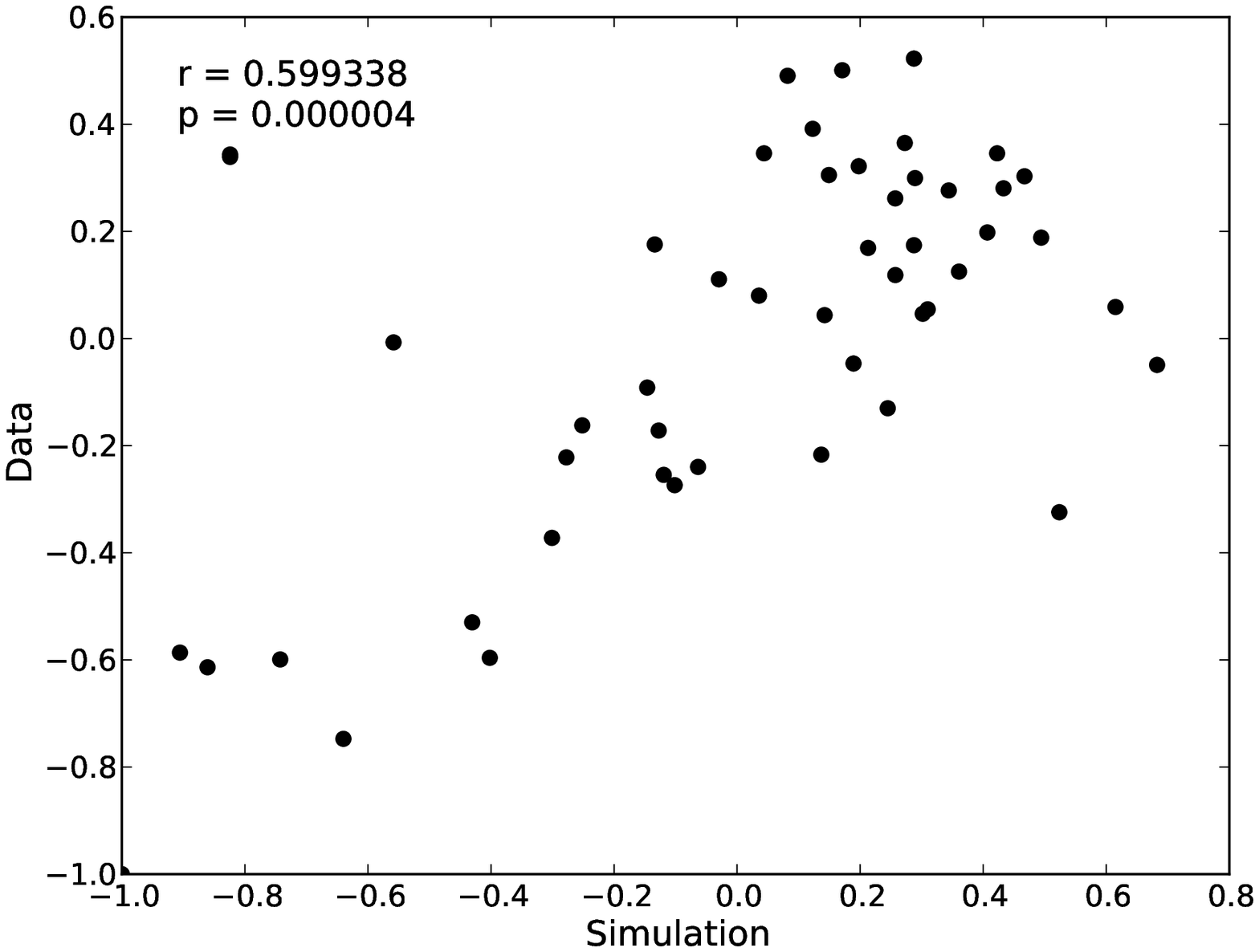}
\caption{\label{fig:xy-plot}
An $xy$-plot of the simulation data ($x$-axis) and observed data
($y$-axis), after substracting the mean and normalization.}
\end{figure}

The resulting particle flux, seen in figure~\ref{fig:accretion}, shows several
accretion peaks induced by the secondary perturbation. The peaks are approximately 
$\sim 4$ years apart and of the same length. 
\rv{An analysis with the Lomb--Scargle periodogram gives the highest
periodic component at 9.26 years, which when scaled back to the physical ISCO
gives a period of 60 days.}
In figure~\ref{fig:phase_diagram} we see a
phase wrapped version of the data in figure~\ref{fig:accretion}.
The outburst profile is very similar to observations. 
\rv{%
We quantify this by calculating the Pearson correlation coefficient of
both the observed and simulated phase wrapped data.
Figure~\ref{fig:xy-plot} shows an $xy$-plot of both the simulated and
observed data. We find the Pearson correlation coefficient to be $r =
0.599$ with a significance of $p=4.25\cdot10^{-6}$, which is clearly
significant.
The theoretical 60 day periodicity is not significantly different from
the observed 50 day period, considering that we have not been able to
simulate the mass flow exactly at ISCO. 
} 

Based on \rv{the theoretical suggestion of the existence of spiral
density waves in a system like \oj{}, and } this simulation we conclude
that the 50 day period is very likely related to accretion in a disc
with a spiral density wave.

\section{Summary}

The binary model of \oj{} was initially constructed in order to understand the
time structure of major outbursts in \oj{}. It led to a very definite solution
which is here investigated further. We now look at the brightness evolution
outside the major outbursts, and look for signatures of the ISCO in the primary
accretion disc. Indeed we find evidence for a 50 day time structure which is an
independent prediction of the model. In addition, the Lorentz contracted time
structure of 3.5 days also shows up.

We have studied the accretion process in a perturbed disc in more detail. We
find that the accretion profile of the 50 day feature agrees well with the
observed average 50 day brightness profile, accumulated over six cycles. 

We may stress the importance of the dense coverage of observations of \oj{}. In
a more sparsely sampled light curve we would not be able to recognize these
temporal features. It is also significant that there are no shorter time-scale
features that appear significant. It would seem to imply that we do not see the
signal from the secondary at this time. It may be too weak, or as in the binary
model, hiding behind the primary disc during the major part of this observing
campaign.

\section*{Acknowledgments}

P.~Pihajoki is supported by the Magnus Ehrnrooth foundation (grant No. Ta2012n6).
We acknowledge funding from European Commission’s Human Potential Programme
(Training and Mobility through Research programme) under contract
HPRN-CT-2002-00321. This work is partly based on data taken and assembled by
the WEBT Collaboration and stored in the WEBT archive at the INAF Observatory
of Torino, Italy (http://www.oato.inaf.it/blazars/webt/), through an intensive
multifrequency campaign. This work is partly based on a ENIGMA (European
Network for the Investigation of Galactic Nuclei through Multifrequency
Analysis) long-term observing campaign. The PI institution and coordinating
partner of the ENIGMA Network is the F\"{o}rderkreis der
Landessternwarte Heidelberg, K\"{o}nigstuhl, Heidelberg, Germany
(http://www.lsw.uni-heidelberg.de/projects/enigma/).

\rv{We are thankful to the anonymous reviewer for comments that have
been very useful in improving this article.}

\bibliography{mn-jour,references}

\bsp

\label{lastpage}

\end{document}